\documentclass[sn-mathphys-ay]{sn-jnl}

\usepackage{graphicx}%
\usepackage{multirow}%
\usepackage{amsmath,amssymb,amsfonts}%
\usepackage{amsthm}%
\usepackage{mathrsfs}%
\usepackage[title]{appendix}%
\usepackage{xcolor}%
\usepackage{textcomp}%
\usepackage{manyfoot}%
\usepackage{booktabs}%
\usepackage{algorithm}%
\usepackage{algorithmicx}%
\usepackage{algpseudocode}%
\usepackage{listings}%
\usepackage{caption}
\usepackage{subcaption}

\theoremstyle{thmstyleone}%
\theoremstyle{thmstyletwo}%

\theoremstyle{thmstylethree}%

\begin{document}

\title[Article Title]{Pair creation in the vortex-driven magnetic fields of black holes}


\author{\fnm{Zaza N.} \sur{Osmanov}}\email{z.osmanov@freeuni.edu.ge}



\affil{\orgdiv{School of Physics}, \orgname{Free University of Tbilisi}, \city{Tbilisi}, \postcode{0183},  \country{Georgia}}

\affil{ \orgname{E. Kharadze Georgian National Astrophysical Observatory}, \city{Abastumani}, \postcode{0301},  \country{Georgia}}


\abstract{We study the effects of pair creation on the radiation emerging from black holes under the assumption that the magnetic fields are vortex driven. In particular, for a sufficiently broad range of supermassive black holes, we investigated the energies at which photons undergo decay under the influence of a strong magnetic field, producing electron-positron pairs. Depending on particular physical parameters, it has been shown that in certain scenarios high or very high energy emission generated by black holes will be strongly suppressed, thus, will be unable to escape a zone where radiation is generated. In particular, photons with energies exceeding $\sim 1$ GeV will never leave the magnetosphere if they are generated at the scale 10$R_g$ and the threshold is of the order of $1$ TeV, if the emission is produced at $\sim 100\; R_g$. Analysing the process versus the black hole mass, assuming the region $100\; R_g$, it has been shown that for the considered lowest mass, the photons with energies $250$ GeV will never leave the black hole and for the considered highest mass the corresponding value is $\sim 250$ TeV.}




\maketitle

\section{Introduction}

It is widely accepted that the processes taking place in active galactic nuclei (AGN) depend on the characteristics of their magnetospheres \citep{ahar,shapiro}. In particular, the physical properties of the magnetosphere are crucial in the study of jet formation \citep{BZ}, particle acceleration \citep{blazar,rieger}, non-thermal radiation \citep{rieger,bodo}, heating of the magnetosphere \citep{heating} and various instabilities \citep{jet,twist}.

In \citep{ruffini} it has been shown that the production of electron-positron pairs is very important in astrophysical processes and may account for the pattern of radiation we observe in the sky. For example, it is believed that the presence of electron-positron pairs might determine as the characteristics of non-thermal radiation \citep{ep1,ep2} and the galactic $\gamma$-ray background \citep{ep3}, as well as the formation of relativistic beams \citep{ep4}.

A dominant pair creation mechanism in the ambient of black holes is $\gamma+\gamma$ process, when relatively low energy photons originating from the accretion disk encounter the high energy photons produced in the photon sea of the accretion disk by means of the inverse Compton scattering \citep{ahar,chen}. In the framework of the Penrose process \citep{penrose}, the accretion disk photons, after reaching the ergosphere and experiencing the blueshift effect, might reach the GeV threshold, which is enough to produce pairs after scattering against protons. We also have shown \citep{PCAGN} that under certain conditions, the centrifugally driven exponentially amplifying electric fields \citep{LangAGN} might reach the Schwinger threshold, $1.4\times 10^{14}$ statvolt cm$^{-1}$ \citep{heisen,sauter,schwinger} with efficient production of electron-positron pairs.

Generally speaking, if the magnetic field, $B$, is extremely high (which takes place in the magnetospheres of pulsars), an additional channel might come into the game, when high energy photons via the reaction $\gamma+B\rightarrow e^{\pm}+B$ decay into pairs. This happens in pulsars where magnetic fields are extremely strong and are of the order of $10^{12}$ G close to the neutron star's surface \citep{GJ,stur,tadem}. In the black hole magnetospheres if the magnetic field is produced by the equipartition scenario (when the emission energy density is of the order of that of the radiation) the value of $B$ might vary in the range $10-10^4$ G  \citep{BZ,rieger}, which is not enough for the realization of the aforementioned process.

Recently, in \citep{vortex,vortex1} a new possibility of generation of magnetic field has been studied. Within an assumption that a black hole is a system of graviton condensate characterised by a vortex structure, which inevitably leads to the trapping of magnetic flux \citep{vortex}. Another independent possibility which also might lead to the same result: trapping of the magnetic field, is an assumption of a nonzero photon mass \citep{chib}. Photon's longitudinal polarization maintains the gauge invariance despite the photon mass \citep{adel} (See also \citep{dvali}), which might originate from a certain extension of the Higgs mechanism  \citep{adel}. It is well known that even for the static scenario, the vector expression ${\bf\nabla}\times{\bf B}$ is not vanishing, but is proportional to the photon mass \citep{vortex} and the corresponding term is a vortex term, which is responsible for generation of the magnetic field. In any of the scenarios the trapped magnetic field is maintained by the black hole itself \citep{vortex} and therefore, its astrophysical context should be significant. In this scenario the maximum value of the magnetic field energy is of the order of the black hole's total energy leading to such high values of magnetic fields that the aforementioned channel $\gamma+B\rightarrow e^{\pm}+B$ becomes significant in the pair production process.

In the paper we study this particular case and for a wide range of black hole masses we explore the effect of pair creation on an observational pattern.

The paper is organized as follows: in Sec.2, we discuss the main aspects of the considered process and applying to black holes we obtain results; and in Sec. 3, we summarize them.

\begin{figure}
\begin{subfigure}[h]{0.55\linewidth}
\includegraphics[width=\linewidth]{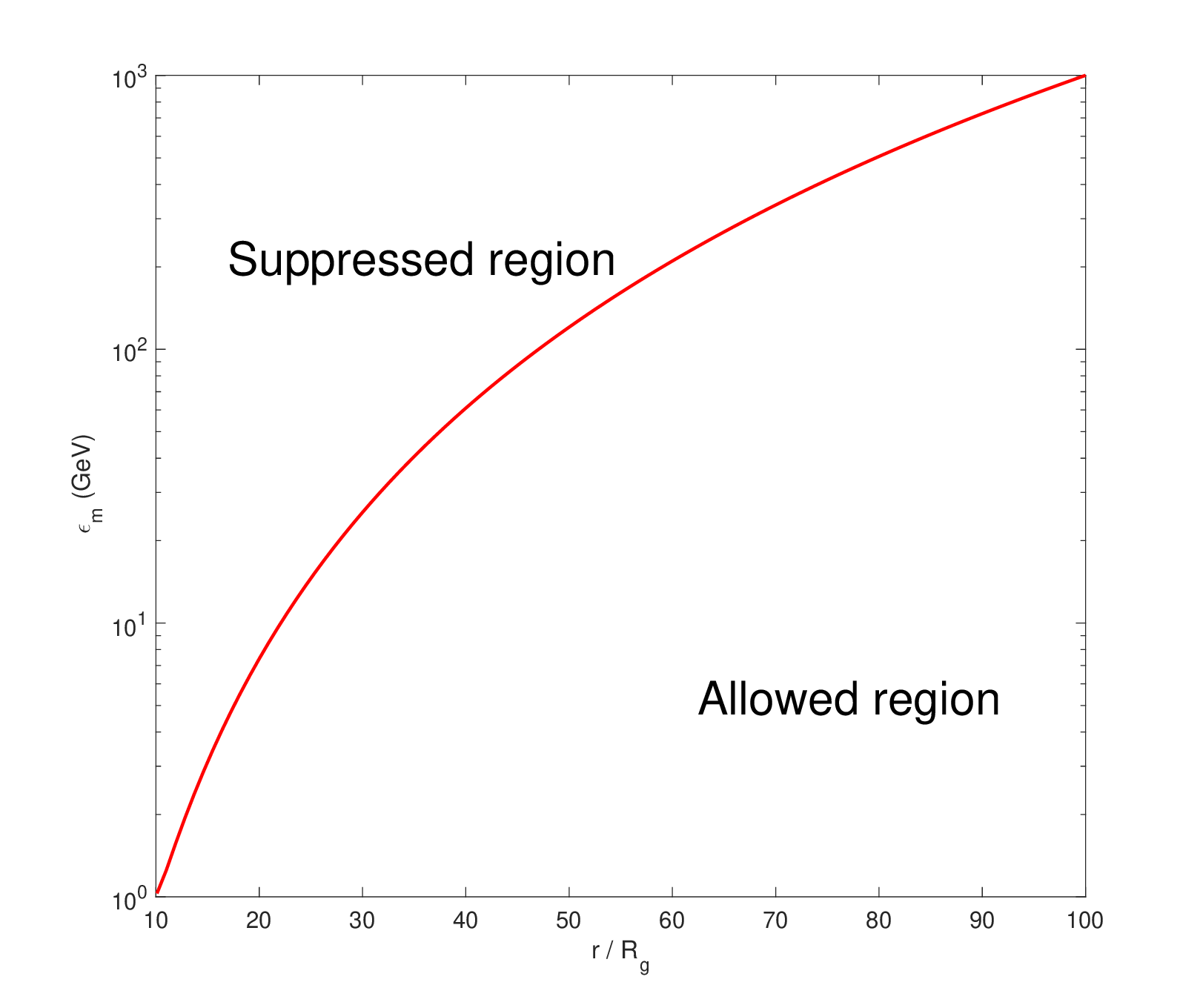}
\end{subfigure}
\hfill
\begin{subfigure}[h]{0.55\linewidth}
\includegraphics[width=\linewidth]{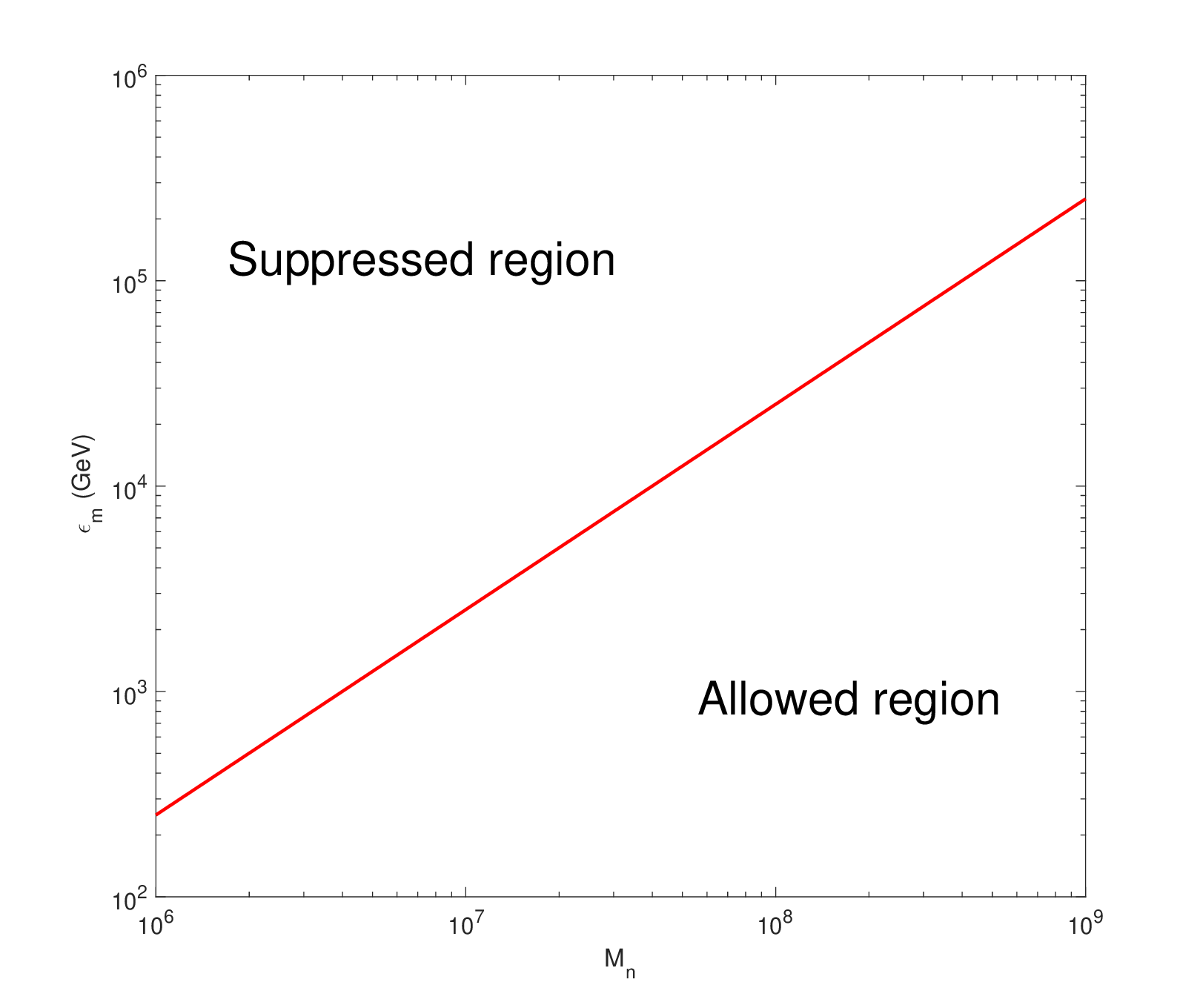}
\end{subfigure}%
\caption{On the left pannel we plot the radial dependence of $\epsilon_m$. The set of parameters is $M = 4\times 10^6 M_{\odot}$, $r_0\simeq 10R_g$, $\theta = 90^0$. On the right pannel we show the graph of $\epsilon_m (M)$. The set of parameters is the same as in Fig. 1a, except $r = 100 R_g$ and $M = (10^6-10^9)\times M_{\odot}$}
\end{figure}

\section{Discussions and results}

In this chapter, we review in detail the aforementioned pair creation channel for magnetospheres of supermassive black holes and estimate the rate of the process and the efficiency population of magnetospheres. 

By considering the rapidly rotating black hole, the corresponding maximal vortex will lead to the condition, when the energy of the maximum magnetic field trapped by the black hole will be of the order of the black hole' energy $B_0^2R_g^3\simeq Mc^2$
\begin{equation}
\label{B}
B_0\simeq \frac{c^4}{MG^{3/2}}\simeq \frac{2.4\times 10^{12}}{M_7}\; G,
\end{equation}
where $R_g = GM/c^2$ represents the gravitational radius, $G$ is the gravitational constant, $M$ denotes the black hole's mass, $M_7 = M/(10^7\times M_{\odot})$, $M_{\odot}\simeq 2\times 10^{33}$ g represents the solar mass and $c$ is the speed of light. As it is evident, the magnetic field is of the same order of magnitude as in pulsars and therefore, pair creation via the channel $\gamma+B\rightarrow e^{\pm}+B$ might be significant.

In general, if the photon energy, $\epsilon$, satisfies the condition $\epsilon\sin\theta\geq 2mc^2$, ($\theta$ is the angle of the photon's direction of propagation versus the magnetic field and $m$ represents electron's mass), it can produce pairs, but the probability of the process strongly depends on the magnetic field. In particular, the corresponding attenuation coefficient (inverse of a length scale, corresponding to a distance of the pair creation) is given by \citep{erber}
\begin{equation}
\label{alpha}
\alpha = \frac{\pi}{137}\frac{1}{\lambda_{_C}}\frac{B(r)\sin\theta}{B_c}T(\chi),
\end{equation}
where $\lambda_{_C}$ is the Compton wave-length,  $B_c\simeq 4.41\times 10^{13}$ G, $\chi = \frac{1}{2}(\epsilon/mc^2)(B/B_c)$,
\begin{equation}
T(\chi) = \begin{cases}
0.46\; \exp\{-4/(3\chi)\} & \quad \chi << 1, \\
0.60\; \chi^{-1/3} & \quad \chi >> 1
\end{cases}
\end{equation}
and we assume the dipolar magnetic field $B(r)\simeq B_0 \;R_g^3/r^3$. It is worth noting that such a behavior in a leading term is a direct consequence of the absence of magnetic monopoles. As we will see later, for the reasonable physical parameters $\chi << 1$.

As an assumption, considering the aforementioned condition, for the location of the absorbing surface (where the photons are absorbed and the pairs are created), one can write $\alpha (r-r_0)\simeq 1$, leading to the expression of minimum photon energy (normalised on the electron's rest energy) when pair creation starts
\begin{equation}
\label{epsilon}
\frac{\epsilon_m}{mc^2}\simeq\frac{8}{3}\frac{B_c}{B_0\sin\theta}\left(\frac{r}{R_g}\right)^3\left[ln\left(\frac{0.46\pi}{137\lambda_{_c}} \frac{B_0\sin\theta}{B_c}\left( \frac{R_g}{r} \right)^3 (r-r_0) \right)\right]^{-1},
\end{equation}
where $r_0$ is the radial coordinate where high energy photons are generated.

As an example, on Fig. 1 (left pannel) we show the radial behaviour of $\epsilon_m (r)$ for the black hole located in our galaxy SgrA* ($M = 4\times 10^6M_{\odot}$ \citep{EHT}). The set of parameters is $M = 4\times 10^6M_{\odot}$, $r_0\simeq 10R_g$, $\theta = 90^0$. From the obtained values one can straightforwardly show that our assumption $\chi<<1$ is satisfied. We have examined the interval $(10-100)\times R_g$ for the radial coordinate, because it is widely accepted that a region where the high energy non-thermal radiation is generated lies in the mentioned area \citep{AGN}. Although, one should note that black holes generate VHE emission not only in the magnetosphere, but also in the jets, far from the central regions \citep{AGN}, where the magnetic field is not as high as in the nearby zones of the black hole. Therefore, we focus only on the aforementioned interval of radial distances. As it is clear, the threshold photon energy is a continuously increasing function of the radial coordinate, which is a natural result, because for larger values of $r$, the magnetic field is smaller and one requires more energetic photons for driving the pair creation. From this plot one can conclude that if the radiation is generated in the inner region ($r\simeq 10R_g$), only photons with energies less than $\sim 1$ GeV can escape the zone. Higher energy photons can escape only if they are generated in the outer regions of the mgnetosphere. In particular, for $r\simeq 100R_g$ only the photons with energies greater than $1$ TeV are suppressed, which means that the corresponding luminosity should be smaller than the bolometric luminosity. In this context it is important to emphasize that the multiwavelength observations, apart from the strong radio and the hard $X$-ray emission, reveal also emission in the TeV domain with the flux $4\times 10^{-12}$ TeV/(cm$^2$ s$^2$) \citep{nature}, which when translated into the corresponding luminosity is much smaller than the bolometric luminosity, $5\times 10^{35}$ erg/s  \citep{nature}, of this object.

On Fig. 1 (right pannel)  we plot the graph $\epsilon_m (M)$. The set of parameters is the same as in Fig. 1, except $r = 100 R_g$ and $M = (10^6-10^9)\times M_{\odot}$. Like the previous case, the minimum energy is an increasing function of the black hole mass, because, as it is clear from Eq. (\ref{B}), the higher the mass the smaller the magnetic field, requiring the higher energies of photons to provoke the pair creation. From this figure one can conclude that emission generated in the outer region of the mgnetosphere cannot escape the zone if the energy exceeds $\sim 250$ GeV ($M = 10^6\times M_{\odot}$) and $\sim 250$ TeV ($M = 10^9\times M_{\odot}$). From the list of the supermassive black holes, one of the interesting examples is $M87$, exceeding the solar mass by several billion times. This object is mostly visible in low spectral energy bands $< 1$ TeV, which in light of the vortex magnetic field could be quite interesting.

It is clear that the photons with energies grater than $\epsilon_m$ will be strongly suppressed by the pair creation process and therefore, one of the important signatures of the existence of the vortex driven magnetic field, could be the absence or at least the presence of strongly suppressed fluxes from the mentioned spectral zone. 

By means of the process $\gamma+B\rightarrow e^{\pm}+B$, the magnetosphere will be seeded by the electron positron pairs, but on the other hand, the increase of the number of pairs will lead to the corresponding increase of the efficiency of the inverse reaction, the annihilation of the pairs with the following rate per unit of volume
\citep{ann} 
\begin{equation}
\label{ann} 
\Lambda\simeq 2\pi c r_e^2 n_{\pm}^2,
\end{equation} 
where $n_{\pm}$ is the number density of electrons and positrons, and $r_e$ is the electron's classical radius. After estimating the energy flux, corresponding to the photon energy $\epsilon$ ($\epsilon\geq\epsilon_m$), $F_{\epsilon}\simeq L_{\epsilon}/(4\pi r^2)$, for the number density of photons one obtains, $n_{ph}\simeq L_{\epsilon}/(4\pi cr^2\epsilon)$. It worth noting that since $\epsilon\geq\epsilon_m$ the conversion rate is almost $100 \%$ and consequently the corresponding emission will not be able to leave the region. We have taken into account that on the scales $10^{1-2}R_g$, the magnetic field is much smaller than $B_c$ and the annihilation rate can be approximated by Eq. (\ref{ann}) \citep{daugh}. By combining the order of magnitude of the escape time-scale, $\delta t\sim r/c$, and the photon number density, one obtains the average pair creation rate per unit of volume:  
\begin{equation}
\label{annR} 
R\simeq \frac{L_{\epsilon}}{4\pi cr^2\epsilon}\times \frac{c}{r}\simeq\frac{L_{\epsilon}}{4\pi r^3\epsilon}.
\end{equation} 
Eventually the process will inevitably saturate, when the pair creation and annihilation rates come to the balance, $R\simeq\Lambda$, and the "equilibrium" number density of electron-positron pairs of the equilibrium state will be established
\begin{equation}
\label{annR} 
n_{\pm}\simeq\left(\frac{L_{\epsilon}}{8\pi^2 r^3r_e^2\epsilon}\right)^{1/2}\simeq 5.5\times 10^9\times\left(\frac{L_{\epsilon}}{10^{42}\;erg/s}\times\frac{GeV}{\epsilon}\right)^{1/2}\; cm^{-3}.
\end{equation} 
Such an increase in the number density of pairs will inevitably lead to the corresponding outflow which must be detectable as additional annihilation lines of the positrons encountering outer electrons and an electron-positron plasma flow, which must be relativistic, because the decaying photons in the GeV domain produce electron-positron pairs with the Lorentz factors, $\gamma\simeq 1 GeV/(2\times 0.511 MeV)\simeq 1000$. 

\section{Conclusions}

Based on an assumption that some black holes might generate the vortex driven magnetic field, we have considered a pair creation channel $\gamma+B \rightarrow e^{\pm}+B$, which only occurs in very strong magnetic fields and normally takes place in pulsars.

Examining the black hole located in the galactic center - SgrA*, and by taking the attenuation coefficient of photons into account, we have found that photons with $\gtrsim 1$ GeV will never leave the black hole zone if they are produced at $\sim 10 R_g$. If on the other hand, the emission is produced in the outer region, $\sim 100 R_g$, only photons with energies exceeding $\sim 1$ TeV will be strongly suppressed.

In the same manner, for a wide variety of supermassive black holes, $10^{6-9}\times M_{\odot}$, we have estimated the minimum energies of the photons, when by means of the aforementioned channel the emission for the corresponding energies is strongly suppressed and does not leave the region where it has been generated. In particular, considering the generation at $\sim 100 R_g$, the threshold energies will be of the order of $250$ GeV ($M = 10^{6}\times M_{\odot}$) and $250$ TeV ($M = 10^{9}\times M_{\odot}$) respectively.

These estimates could be significant to identify the origin of the magnetic field - it is of vortex-driven or not. In particular, the absence of certain high energy emission spectral lines and the presence of outflows with additional annihilation lines should be a direct indication of the vortex origin of magnetic field of the black hole.

On the other hand, we should note that there is no direct method for determining the magnetic field of a black hole, so the only way is to look for it indirectly. It is known that the energy of some particles coming from space is very high, and the mechanisms of their generation are still under question. For example, the discovery of ultra-high energy particles of the order of $10^{20}$ eV \citep{amaterasu,2007,1993,1991} seriously question existing models of particle acceleration, although the vortex-driven magnetic field can guarantee extremely high energies. In particular, the Hillas criterion postulates that the maximum achievable energy of particles should satisfy the condition $E\leq eBR_g$, and the upper limit for the vortex driven magnetic field is of the order of $10^{27}$ eV.

\section*{Acknowledgements}

The research was supported by Shota Rustaveli National Science Foundation of Georgia (SRNSFG) Grant: FR-23-18821 and a German DAAD scholarship within the program Research Stays for University Academics and Scientists, 2024 (ID: 57693448). ZO is grateful to prof. G. Dvali for fruitful discussions and comments. ZO acknowledges Max Planck Institute for Physics (Munich) for hospitality during the completion of this project.
Please refer to Journal-level guidance for any specific requirements.

\section*{Data Availability}
No datasets were generated or analysed during the current study.

\section*{Funding}
Funding: The research was supported by Shota Rustaveli National Science Foundation of Georgia (SRNSFG) Grant: FR-23-18821 and a German DAAD scholarship within the program Research Stays for University Academics and Scientists, 2024 (ID: 57693448). 
\section*{Ethics declaration}
Not applicable.

\end{document}